\documentclass[a4paper,11pt]{article}
\pdfoutput=1 

\usepackage{jheppub} 
\usepackage{amssymb,amsmath,verbatim,mathtools,needspace,enumitem,etoolbox,graphicx,physics,microtype,afterpage,xspace,tabularx,lmodern,multirow,boldline, makecell, booktabs}

\usepackage{appendix}
\usepackage{tensor}
\usepackage{array}
\usepackage{siunitx}
\usepackage[normalem]{ulem}
\usepackage[dvipsnames, usenames]{xcolor}
\usepackage{xr-hyper}
\usepackage[font=small,labelfont=bf]{caption}
\definecolor{linkcolor}{rgb}{0.0,0.3,0.5}
\usepackage[all]{hypcap}
\usepackage[T1]{fontenc}
\usepackage[utf8]{inputenc}
\usepackage[usenames,dvipsnames]{xcolor}
\usepackage[T1]{fontenc} 
\usepackage{tikz-cd}
\usepackage{caption}
\usepackage{xcolor}
\title{\boldmath  Exact  $T\bar{T}$-Like  flows in Integrable Sigma Models}

\author[a]{H. Babaei-Aghbolagh,}

\affiliation[a]{Institute of Fundamental Physics and Quantum Technology, \& School of Physical Science and Technology, Ningbo University, Ningbo, Zhejiang 315211, China}

\emailAdd{hosseinbabaei@nbu.edu.cn}

\abstract{
Recently, a general exact autonomy framework for $T\bar{T}$-like deformations of 2D integrable sigma models and 4D nonlinear electrodynamics has been developed. The theories are characterized by the Courant–Hilbert equation and admit an equivalent description in the Russo–Townsend auxiliary-field formulation. For general one-scale families, we derive an exact irrelevant flow equation of the universal form
$
\partial_\lambda \mathcal L
=X\mathcal F(\sqrt{Y/X}),
\qquad
X=T_{\mu\nu}T^{\mu\nu},
\qquad
Y=(T^\mu{}_\mu)^2,
$
where the function $\mathcal F$ is determined by the underlying Courant–Hilbert solution. We obtain a parametric representation valid for arbitrary one-scale models, establish its perturbative expansion in terms of the defining function or auxiliary potential, and identify the domain in which the flow remains regular. We also show that the irrelevant flow commutes with the marginal root-$T\bar T$ deformation and that both flows preserve classical integrability. Explicit results are derived for generalized Born–Infeld, q-deformed, no-$\tau$-maximum, and logarithmic models, with the latter admitting a closed representation. Finally, we extend the flow construction to arbitrary spacetime dimension $(d)$, revealing a common one-variable structure that connects two-dimensional integrable sigma models to four-dimensional self-dual nonlinear electrodynamics. These results provide a systematic framework for constructing and classifying exact $T\bar T$-like flows beyond the standard Born–Infeld setting.}

\begin{document}
\maketitle
\flushbottom

\section{Introduction and  summary of results}
\label{sec:intro}
The $T\bar T$ deformation~\cite{Zamolodchikov:2004ce,Smirnov:2016lqw,Cavaglia:2016oda} is one of the few examples of an irrelevant deformation of a two-dimensional quantum field theory that is under exact control (see~\cite{Jiang:2019epa,He:2025ppz} for reviews). It is defined by the flow
\begin{equation}
\frac{\partial\mathcal L}{\partial\lambda}=\mathcal O_{T\bar T},\qquad
\mathcal O_{T\bar T}\propto\det T_{\mu\nu}\propto X-Y ,
\end{equation}
with $X=T_{\mu\nu}T^{\mu\nu}$ and $Y=(T^\mu{}_\mu)^2$, and it preserves the integrability and solvability of the deformed theory~\cite{Smirnov:2016lqw,Cavaglia:2016oda,Conti:2018jho,Rosenhaus:2019utc,Chen:2021aid}. It has a random-geometry and dynamical-coordinate interpretation~\cite{Cardy:2018sdv,Bonelli:2018kik}, a holographic interpretation as a finite-cutoff boundary condition~\cite{McGough:2016lol,Hartman:2018tkw}, and it admits generalisations with higher-spin currents~\cite{Camilo:2021gro}, higher-dimensional analogues~\cite{Taylor:2018xcy,Conti:2022egv,Brizio:2026ynw} and supersymmetric extensions~\cite{Ferko:2019oyv}. A marginal companion, the root-$T\bar T$ flow generated by $\sqrt{2X-Y}$ in two dimensions, was introduced in~\cite{Ferko:2022cix,Babaei-Aghbolagh:2022leo}; its classical integrability was established in~\cite{Borsato:2022tmu}, and it has a geometric formulation and an algebraic origin in the conformal algebra~\cite{Babaei-Aghbolagh:2024hti,Tempo:2022ndz,Banerjee:2026qyc}. Generalised $T\bar T$-like flows have also been constructed for two-dimensional scalar theories~\cite{Babaei-Aghbolagh:2025lko}, Yang--Mills theory~\cite{Ferko:2024yua} and four-dimensional Chern--Simons theory~\cite{Sakamoto:2025hwi,Fukushima:2025tlj}.

In four dimensions, flows driven by quadratic invariants of the stress tensor generate Born--Infeld~\cite{Born:1934gh,Tseytlin:1999dj,Gibbons:2000xe} and ModMax~\cite{Bandos:2020jsw,Bandos:2021jhn} electrodynamics and their generalisations~\cite{Babaei-Aghbolagh:2020kjg,Babaei-Aghbolagh:2022uij,Ferko:2022iru,Ferko:2023ruw,Ferko:2023wyi,Babaei-Aghbolagh:2024uqp,Babaei-Aghbolagh:2025cni}. Duality-invariant nonlinear electrodynamics~\cite{Gaillard:1981rj,Gibbons:1995cv,Gaillard:1997zr,Bialynicki-Birula:1984daz,Kosyakov:2020wxv,Sorokin:2021tge} is described by a Lagrangian that depends on two variables and satisfies a single nonlinear partial differential equation. This equation has been solved with the help of auxiliary fields~\cite{Ivanov:2002ab,Ivanov:2003uj,Ivanov:2012bq}, and in the causal case by Russo-Townsend~\cite{Russo:2024llm,Russo:2024xnh,Russo:2024ptw,Russo:2025fuc}, using the method of Courant-Hilbert~\cite{Courant:Hilbert}. The Courant--Hilbert (CH) solution is connected to the method of characteristics~\cite{Hou:2022csf,Chen:2025ndc}, has been generalised in~\cite{Kuzenko:2025gvn,Kuzenko:2026hbw}, and has been applied in four-dimensional Chern--Simons theory, in Yang--Baxter sigma models, and in gravity and black-hole physics~\cite{Fukushima:2025tlj,Fukushima:2026gan,Fukushima:2026znr,Babaei-Aghbolagh:2025tim,Adami:2025pqr,Adami:2025pfk, Adami:2026jmo}.

The operators in the examples above are polynomials, or square roots, of $X$ and $Y$. The generic situation is different. If a two-dimensional field theory is integrable, one would like to know which operator, built from its own stress tensor, generates it from the conformal point, and whether that operator stays under control far from the perturbative regime. Starting from a theory with Lagrangian density $\mathcal L_{\rm CFT}$ and a dimensionful coupling $\lambda$, any deformed theory can formally be written as
\begin{equation}
\mathcal L=\mathcal L_{\rm CFT}+\int\mathcal O_\lambda(T_{\mu\nu})\,d\lambda ,
\end{equation}
where $\mathcal O_\lambda$ is an irrelevant operator whose leading term is $\mathcal O(T_{\mu\nu}T^{\mu\nu})$. Its leading behavior is the Zamolodchikov operator, but beyond leading order $\mathcal{O}_\lambda$ is not known for integrable models other than the Born-Infeld-like case. Determining it exactly and nonperturbatively for a broad class of integrable theories is the problem addressed in Ref.~\cite{Babaei-Aghbolagh:2026gbr}.\footnote{This question was posed for integrable theories by Zamolodchikov in his fourth lecture on $T\bar T$ deformations of two-dimensional quantum field theories at the 36th Advanced School in Physics, \emph{Recent Progress in Quantum Field/String Theory} (30 December--10 January 2019).}

Integrable sigma models provide a natural arena. The Lagrangians of integrable sigma models built on the principal chiral model (PCM)~\cite{Gell-Mann:1960mvl} that depend on the two currents are characterised by a single nonlinear Courant–Hilbert  equation, in the two variables  constructed from them, and are all compatible with the Lax connection of the PCM~\cite{Ferko:2024ali,Bielli:2024ach}. This is the family of auxiliary field sigma models (AFSMs), which has been extended to $T$-duality and $T\bar T$-like deformations~\cite{Bielli:2024khq}, Yang--Baxter deformations~\cite{Bielli:2024fnp}, cosets and dimensionally reduced gravity~\cite{Cesaro:2024ipq,Cesaro:2025msv,Bielli:2026vit,Baglioni:2026mii}, the $\mathbb{CP}^{N-1}$ and lambda models~\cite{Ferko:2025bhv,Ferko:2026eci}, higher-spin currents~\cite{Bielli:2025uiv} and four-dimensional Chern--Simons theory~\cite{Fukushima:2024nxm}. The Courant–Hilbert equation  is the two-dimensional counterpart of the equation satisfied by causal self-dual nonlinear electrodynamics. The relation between the four-dimensional and the two-dimensional auxiliary field formulations was worked out in~\cite{Baglioni:2025tsc}, and the Courant--Hilbert approach was transferred to two dimensions in~\cite{Babaei-Aghbolagh:2025uoz,Babaei-Aghbolagh:2025hlm}. The solution is determined by one function of one variable, the CH function $\ell(\tau)$; equivalently, by a potential $\Omega$ in the Legendre-transformed Russo--Townsend description, which reproduces the interaction function of the AFSMs. The root-$T\bar T$ deformation of the PCM is a one-parameter family inside this class, and the $T\bar T$-like operator of a model with $\ell$ expanded around the PCM point,
\begin{equation}\label{ell}
\ell(\tau)=e^\gamma\tau+a_1\lambda e^{2\gamma}\tau^2+a_2\lambda^2e^{3\gamma}\tau^3+\cdots,
\end{equation}
is the object we compute.

\paragraph{Summary of results.}
\begin{itemize}
\item \emph{Exact $T\bar T$-like flow} (sections~\ref{sec:CH} and~\ref{sec:RT}). For one-scale solutions $\ell=\lambda^{-1}f(e^\gamma\lambda\tau)$ the stress tensor depends on $\tau\ell'$ and $\ell-\tau\ell'$ only, so $X$ and $Y$ are functions of one variable, and eliminating it gives
\begin{equation}
\frac{\partial\mathcal L}{\partial\lambda}=X\,\mathcal F(\rho),\qquad\rho=\sqrt{Y/X},\qquad
\mathcal F(\rho)=\sum_{m\ge0}C_m\rho^m,
\label{introflow}
\end{equation}
together with the universal marginal flow $\partial_\gamma\mathcal L=\tfrac12\sqrt{2X-Y}$. The right-hand side contains no explicit dependence on $\lambda$, $\gamma$ or the fields. The coefficients $C_m$ are given in terms of the coefficients of $\ell$, or of the potential: $C_0=a_1/2$, $C_1=(a_2-2a_1^2)/(\sqrt2a_1)$; with the potential $\Omega=\lambda^{-1}(\omega_2\phi^2+\omega_3\phi^3+\cdots)$, $C_0=1/(8\omega_2)$ and $C_1=-\omega_3/(2\sqrt2\omega_2^2)$. Integer powers of $X$ and $Y$ occur if and only if all odd $C_m$ vanish.
\item \emph{Causality and singular points} (section~\ref{sec:stress}). The causal window $\ell'\ge1$, $\ell''\ge0$ is expressed through the eigenvalues of the stress tensor, and fixes the physical branch and the range of $\rho$; each endpoint is identified as a critical current or as a strong-field limit.
\item \emph{Four exact examples} (sections~\ref{sec:GBI}--\ref{sec:log}). The generalised Born--Infeld model, where the flow is the $T\bar T$ operator itself; the $q$-deformed family, which interpolates between ModMax and the logarithmic model; a model without maximal field strength, in closed algebraic form; and the logarithmic model, whose flow involves the Lambert $W$ function.
\item \emph{Commutativity and integrability} (sections~\ref{sec:AFSM} and~\ref{sec:comm}). The irrelevant and the root-$T\bar T$ flows commute exactly, and, since every model shares the Lax connection of the PCM, all flows preserve classical integrability.
\item \emph{Arbitrary dimension} (section~\ref{sec:anyd}). The flow is $\partial_\lambda\mathcal L=\frac{dX-Y}{d^2}\Psi(r)$ with $r=-T^\mu{}_\mu/\sqrt{dX-Y}$ and a $d$-independent function $\Psi$. For $d=2$ it is the sigma-model flow above, and for $d=4$ the flow of self-dual nonlinear electrodynamics~\cite{Babaei-Aghbolagh:2025cni,Pezhman:2026dsr,Pezhman:2026sby,Pezhman:2026gbv}; coefficients scale as $C_m^{(d)}=(2/d)^{1+m/2}C_m^{\rm 2D}$.
\end{itemize}

The paper is organised as follows. In section~\ref{2.0} we review integrable sigma models and the equation $\mathcal L_u\mathcal L_v=-1$. Section~\ref{sec:CH} derives the exact $T\bar T$-like flows from the Courant--Hilbert solution and analyses the stress tensor and causality, section~\ref{sec:RT} gives the Russo--Townsend formulation, and section~\ref{sec:AFSM} its relation to auxiliary field sigma models and to integrability. Sections~\ref{sec:GBI}--\ref{sec:log} treat the four examples. Section~\ref{sec:anyd} extends the flow to arbitrary dimension, and section~\ref{sec:comm} proves that the irrelevant and root-$T\bar T$ flows commute. We summarise and discuss open directions in section~\ref{sec:conclusions}.

\section{ Integrable Sigma Models}\label{2.0}

Integrability-preserving deformations of two-dimensional sigma models,
motivated by integrable structures on the string worldsheet, play a
prominent role throughout string theory.

We consider classical field theories on a flat worldsheet $\Sigma$ with
coordinates $\sigma^\pm=\tau\pm\sigma$ and target space a Lie group $G$
with Lie algebra $\mathfrak g$. The field $g(\sigma,\tau)\colon\Sigma\to G$
defines the Maurer--Cartan forms $j=g^{-1}dg$, $\tilde j=-(dg)g^{-1}$,
obeying the flatness condition
\begin{equation}
    \partial_+ j_- - \partial_- j_+ + [j_+,j_-] = 0
    = \partial_+ \tilde j_- - \partial_- \tilde j_+ + [\tilde j_+,\tilde j_-].
\end{equation}
The simplest example is the Principal Chiral Model~\cite{Gell-Mann:1960mvl},
$\mathcal L_{\text{PCM}} = -\tfrac{1}{2}\tr[j_+j_-]$.
For Lagrangians depending on $j_\pm$ only through
\begin{equation}
    P_1 = -\tr[j_+ j_-], \quad
    P_2 = \frac{1}{2}\Big(\tr[j_+ j_+]\,\tr[j_- j_-] + \big(\tr[j_+ j_-]\big)^2\Big),
\end{equation}
the equations of motion take the form $\partial^\mu\mathfrak J_\mu=0$,
$\mathfrak J_\mu = 2\mathcal L_{P_1}j_\mu + 4\mathcal L_{P_2}g^{\nu\rho}\tr[j_\mu j_\nu]j_\rho$,
where $\mathcal L_{P_i}\equiv\partial\mathcal L/\partial P_i$.
For a deformed theory $\mathcal L(P_1,P_2)$,
integrability requires flatness of the Lax connection~\cite{Faddeev:1987ph,Babelon:2003qt}
$\mathfrak L_\pm = (j_\pm \pm z\mathfrak J_\pm)/(1-z^2)$
for all spectral parameters $z$, equivalent to~\cite{Borsato:2022tmu}
\begin{equation}
    8(P_1^2-P_2)\mathcal L_{P_2}^2 + 8P_1\mathcal L_{P_1}\mathcal L_{P_2} + 4\mathcal L_{P_1}^2 = 1.
\end{equation}
In terms of
$u,v = \tfrac14\big(\sqrt{2P_2-P_1^2}\mp P_1\big)$,
this reduces to~\cite{Babaei-Aghbolagh:2025uoz}
\begin{equation}\label{2DPDE}
    \frac{\partial\mathcal L}{\partial u}\,
    \frac{\partial\mathcal L}{\partial v} = -1,
\end{equation}
 which governs all integrable deformations of the PCM in this class, including the two-dimensional ModMax theory, the $q$-deformed family, and logarithmic models.~\cite{Babaei-Aghbolagh:2025uoz,Babaei-Aghbolagh:2025hlm}. Various approaches have been proposed to solve this partial differential equation, including the Courant-Hilbert approach~\cite{Courant:Hilbert,Russo:2024llm}, the Ivanov-Zupnik formalism~\cite{Ivanov:2002ab,Ivanov:2003uj,Ivanov:2012bq}, the Russo-Townsend auxiliary field approach~\cite{Russo:2024llm,Russo:2025fuc,Kuzenko:2025gvn}, and constructions based on the Born-Infeld seed theory~\cite{Kuzenko:2026hbw}. As shown in \cite{Baglioni:2025tsc}, these approaches are related by well-known conventions, since all of them ultimately express the theory in terms of two variables, $u$ and $v$. We adopt the Courant-Hilbert approach and then turn to the Russo-Townsend formulation to obtain the corresponding results.

\medskip
\noindent
The two-dimensional problem has a feature that makes it a particularly clean laboratory for
the construction above. The variables $u,v$ are non-negative: by the Cauchy--Schwarz inequality
$(\tr[j_+j_-])^2\le\tr[j_+j_+]\tr[j_-j_-]$, so that $P_1^2\le 2P_2-P_1^2$ and hence $u,v\ge0$. They vanish
simultaneously only at $j_\pm=0$, while $v-u=P_1/2$ is the PCM Lagrangian, $\mathcal L_{\rm PCM}=v-u$.
The Poisson structure of this Lax connection is governed by the Maillet bracket~\cite{Maillet:1985ek,Maillet:1985ec}. The Lax connection above is the same one that appears for the auxiliary field sigma models of
Ferko and Smith~\cite{Ferko:2024ali}, and for the higher-spin generalisations of~\cite{Bielli:2024ach,Bielli:2025uiv};
the common twist function is $\varphi(z)=(z^2-1)/z^2$ \cite{Bielli:2024ach}. Consequently every solution of
\eqref{2DPDE} is classically integrable, and in particular so is every theory generated by the
construction of the next section.
The self-duality equation \eqref{2DPDE} is, word for word, the equation $\mathcal L_U\mathcal L_V=-1$
that characterises duality-invariant nonlinear electrodynamics in four dimensions, with
$(U,V)\to(u,v)$; this is the origin of the dictionary between the two problems
used in \cite{Babaei-Aghbolagh:2025uoz,Babaei-Aghbolagh:2025hlm,Baglioni:2025tsc} and is the reason why the
Courant--Hilbert (CH) technology of Russo-Townsend \cite{Russo:2024llm,Russo:2024xnh} can
be transplanted to two dimensions.

\section{Exact Deformations: Courant--Hilbert approach}\label{sec:CH}
\subsection{Courant--Hilbert solution and the stress tensor}

The general solution of \eqref{2DPDE} is parametrised by a single function $\ell(\tau)$ of one
variable~\cite{Babaei-Aghbolagh:2025hlm},
\begin{equation}
\mathcal{L}
=\ell(\tau)-\frac{2u}{\ell'(\tau)},
\qquad
\tau=v+\frac{u}{[\ell'(\tau)]^{2}} ,
\label{CHsol}
\end{equation}
where $\ell'\equiv d\ell/d\tau\neq0$ and the physical domain is $\tau\ge0$, with $\tau=0$ at $u=v=0$.
Differentiating \eqref{CHsol} gives $\mathcal L_v=\ell'$ and $\mathcal L_u=-1/\ell'$, so that \eqref{2DPDE}
holds identically. The Principal Chiral Model corresponds to $\ell(\tau)=\tau$: then $\tau=u+v$ and
$\mathcal L=v-u$. The two-dimensional ModMax theory, i.e.\ the root-$T\bar T$ deformed
PCM~\cite{Ferko:2022cix,Babaei-Aghbolagh:2022leo,Borsato:2022tmu}, is $\ell(\tau)=e^{\gamma}\tau$, for which
$\mathcal L=e^{\gamma}v-e^{-\gamma}u$. Following the four-dimensional analysis of Russo and
Townsend~\cite{Russo:2024llm,Russo:2024xnh}, we normalise $\ell(0)=0$ and call a theory \emph{causal}
(or convex) if
\begin{equation}
\ell'\ge1,\qquad \ell''\ge0 .
\label{causal}
\end{equation}
These inequalities will select the physical branch of every example below.

The energy--momentum tensor of a theory $\mathcal L(P_1,P_2)$ depends on $\ell$ only through the
two combinations $\tau\ell'$ and $\ell-\tau\ell'$. Defining the Hilbert stress tensor as usual~\cite{Babaei-Aghbolagh:2025hlm},
\begin{equation}
T_{\mu\nu}
=-\,\frac{\tau\ell'}{u+v}\,
T_{\mu\nu}^{\mathrm{PCM}}
+(\ell-\tau\ell')\,g_{\mu\nu},
\qquad
T_{\mu}{}^{\mu}=2\left(\ell-\tau\ell'\right),
\label{Tdecomp}
\end{equation}
where $T^{\rm PCM}_{\mu\nu}$ is the traceless PCM stress tensor, normalised such that
$T^{\rm PCM}_{\mu\rho}T^{{\rm PCM}\,\rho}{}_{\nu}=(u+v)^2g_{\mu\nu}$. The trace follows independently
from the scaling argument $T^\mu{}_\mu=2(\mathcal L-u\mathcal L_u-v\mathcal L_v)$, since $u,v$ are homogeneous of
degree one in the inverse worldsheet metric, together with $u\mathcal L_u+v\mathcal L_v=\tau\ell'-2u/\ell'$.
The two quadratic invariants
\begin{equation}
\begin{aligned}
X\equiv T_{\mu\nu}T^{\mu\nu}
&=2\left[(\tau\ell')^{2}+(\ell-\tau\ell')^{2}\right],
\\
Y\equiv T_{\mu}{}^{\mu}T_{\nu}{}^{\nu}
&=4(\ell-\tau\ell')^{2},
\end{aligned}
\label{XY}
\end{equation}
are therefore functions of $\ell$ and $\tau\ell'$ only. In two dimensions
$\det T^\mu{}_\nu=\tfrac12(Y-X)$, so that the Zamolodchikov $T\bar T$ operator is a particular
linear combination of $X$ and $Y$. We also record the traceless part, which has
$\hat T_{\mu\nu}\hat T^{\mu\nu}=2(\tau\ell')^2=\tfrac12(2X-Y)$ and thus
\begin{equation}
\tau\ell'=\tfrac12\sqrt{2X-Y}\qquad(\tau\ell'>0).
\label{rootop}
\end{equation}
This is the root-$T\bar T$ operator~\cite{Ferko:2022cix,Babaei-Aghbolagh:2022leo}, $\mathcal R_{\sqrt{T\bar T}}=\tfrac12\sqrt{2X-Y}$.

\subsection{Deformation parameters}

For any parameter $\alpha$ entering $\ell$ (the coupling $\lambda$, the marginal parameter
$\gamma$, \dots), the dependence of the Lagrangian on $\alpha$ can be extracted without solving the
constraint for $\tau(u,v)$,
\begin{equation}
\left.
\frac{\partial \mathcal{L}}{\partial \alpha}
\right|_{u,v}
=
\left.
\frac{\partial \ell}{\partial \alpha}
\right|_{\tau}.
\label{id}
\end{equation}

\paragraph{Lemma 1.} Equation \eqref{id} holds for every parameter $\alpha$ of $\ell$.

\emph{Proof.} Write $h=\ell'$. At fixed $u,v$ the constraint in \eqref{CHsol} gives
$(1+2uh_\tau/h^3)\,\delta\tau=-2u\,h_\alpha\,\delta\alpha/h^3$. Varying \eqref{CHsol},
$\delta\mathcal L=\ell_\alpha\delta\alpha+h\,(1+2uh_\tau/h^3)\,\delta\tau+2uh_\alpha\delta\alpha/h^2=\ell_\alpha\delta\alpha$. $\square$

This is dimension independent: the same statement holds for the four-dimensional
self-dual electrodynamics with $(U,V)$ in place of $(u,v)$.

\subsection{One-scale solutions and the exact $T\bar T$-like flows}

We consider theories with one dimensionful coupling $\lambda$ (of mass dimension $[\lambda]=-2$,
since $[\mathcal L]=[\ell]=[\tau]=2$ in two dimensions) and one dimensionless parameter $\gamma$.
The most general such Courant--Hilbert function is
\begin{equation}
\ell(\tau;\lambda,\gamma)
=
\frac{1}{\lambda}f(\zeta),
\qquad
\zeta=e^{\gamma}\lambda\tau ,
\label{oneScale}
\end{equation}
with the normalisation $f(0)=0$, $f'(0)=1$, so that $\ell=e^\gamma\tau+\mathcal O(\lambda)$.
The expansion \eqref{ell} of the Introduction corresponds to $f(\zeta)=\zeta+a_1\zeta^2+a_2\zeta^3+\cdots$
at $\gamma=0$ ($a_0=1$ after the trivial rescaling of $\tau$). No analyticity of $f$ is needed below.
Define
\begin{equation}
a(\zeta)=\zeta f'(\zeta),
\qquad
b(\zeta)=f(\zeta)-\zeta f'(\zeta).
\label{ab}
\end{equation}
Then
\begin{equation}
\tau\ell'(\tau)=\frac{a(\zeta)}{\lambda},
\qquad
\ell(\tau)-\tau\ell'(\tau)=\frac{b(\zeta)}{\lambda},
\end{equation}
and, by \eqref{id} and \eqref{XY},
\begin{equation}
\begin{aligned}
\frac{\partial\mathcal{L}}{\partial\lambda}
&=-\frac{b(\zeta)}{\lambda^2},
&\qquad
\frac{\partial\mathcal{L}}{\partial\gamma}
&=\frac{a(\zeta)}{\lambda}=\tau\ell',\\[4pt]
X
&=\frac{2\left[a(\zeta)^2+b(\zeta)^2\right]}{\lambda^2},
&
Y
&=\frac{4 b(\zeta)^2}{\lambda^2}.
\end{aligned}
\label{flowsab}
\end{equation}
The marginal flow is therefore universal,
\begin{equation}
\ \frac{\partial\mathcal L}{\partial\gamma}=\tau\ell'=\tfrac12\sqrt{2X-Y}\ ,
\label{gammaflow}
\end{equation}
i.e.\ the root-$T\bar T$ flow, for every one-scale solution~\cite{Babaei-Aghbolagh:2025hlm}. The single trace flow
is,
\begin{equation}
\frac{\partial \mathcal{L}}{\partial \lambda}
=
-\frac{1}{2\lambda}\,T^\mu{}_\mu ,
\end{equation}
which is exact but not autonomous because of the explicit $\lambda$. An autonomous
representation follows by dividing by $X$:
\begin{equation}
\frac{\partial\mathcal{L}}{\partial\lambda}
=
X\,F(\zeta),
\qquad
F(\zeta)
=
-\frac{b(\zeta)}
{2\left[a(\zeta)^2+b(\zeta)^2\right]}\ ,
\label{Fzeta}
\end{equation}
with
\begin{equation}
\rho^2\equiv\frac{Y}{X}=\frac{2\,b(\zeta)^2}{a(\zeta)^2+b(\zeta)^2}.
\end{equation}
Equation \eqref{Fzeta} is the exact, non-perturbative parametric $T\bar T$-like flow of every one-scale
integrable sigma model; the coupling enters only through the auxiliary parameter $\zeta$. At the
conformal point $F(0)=f''(0)/4$. The dimensions are consistent: $[\partial_\lambda\mathcal L]=4=[X]=[Y]$.

\subsection{The autonomous flow and its operator content}

Suppose that on a physical interval $I$ the map $\zeta\mapsto\rho^2(\zeta)=Y/X$ is one-to-one. It then
has a branch inverse $\zeta=Z(\rho^2)$, and defining $\mathcal F(\rho)\equiv F(Z(\rho^2))$ we obtain
\begin{equation}
\frac{\partial\mathcal L}{\partial\lambda}=X\,\mathcal F\!\left(\rho\right),\qquad \rho=\sqrt{Y/X}\ .
\label{autonomous}
\end{equation}
The right-hand side contains no explicit dependence on $\lambda$, $\gamma$, $\tau$, $u$ or $v$; it is a
function of the stress-tensor invariants alone. It is convenient to trade $\zeta$ for the signed ratio
\begin{equation}
\begin{aligned}
r(\zeta)
&\equiv-\frac{b(\zeta)}{a(\zeta)}
=-\frac{T^\mu{}_\mu}{\sqrt{2X-Y}},
\\
\rho^2
&=\frac{2r^2}{1+r^2},
\qquad
r=-\operatorname{sgn}(T^\mu{}_\mu)\,
\frac{\rho}{\sqrt{2-\rho^2}},
\\
F(\zeta)
&=\frac{r(\zeta)}
{2a(\zeta)\left[1+r(\zeta)^2\right]}.
\end{aligned}
\label{rrho}
\end{equation}
If $r(\zeta)$ has turning points, the autonomous flow is defined branchwise on the corresponding
intervals of monotonicity. Expanding around the $\rho \to 0$,
\begin{equation}
\mathcal F(\rho)=\sum_{m=0}^{\infty}C_m\rho^m,
\qquad
\frac{\partial\mathcal L}{\partial\lambda}
=
\sum_{m=0}^{\infty}
C_m\,Y^{m/2}X^{1-m/2}.
\label{Cm}
\end{equation}
Even $m$ give integer powers of $X$ and $Y$, odd $m$ half-integer ones. Hence the flow is built from
integer powers of the invariants if and only if all odd coefficients vanish.

The first coefficients are determined by the first Taylor coefficients of $f$. With
$f=\zeta+a_1\zeta^2+a_2\zeta^3+a_3\zeta^4+a_4\zeta^5+\cdots$ one finds
\begin{equation}
\begin{aligned}
C_0&=\frac{a_1}{2},\qquad
C_1=\frac{a_2-2a_1^2}{\sqrt2\,a_1},\qquad
C_2=- \frac{a_2 (a_1^2 + 2 a_2)}{2 a_1^3} + \frac{3 (a_1^3 + a_3)}{4 a_1^2}\\
C_3&=\frac{(2 a_1^2 - 9 a_2) a_3}{2 \sqrt{2} a_1^4} + \frac{2 a_1^6 -  a_1^4 a_2 + 16 a_2^3 + a_1^2 (-6 a_2^2 + 4 a_4)}{4 \sqrt{2} a_1^5}
\end{aligned}
\label{C012}
\end{equation}
In particular the odd coefficient $C_1$ vanishes iff $a_2=2a_1^2$, which is the condition satisfied by the
Born--Infeld-type solution of section~\ref{sec:GBI}.

\subsection{Eigenvalues of $T^\mu{}_\nu$ and the causal window}\label{sec:stress}

Write $a\equiv\tau\ell'$ and $b\equiv\ell-\tau\ell'$ for the unscaled combinations
($a=a(\zeta)/\lambda$, $b=b(\zeta)/\lambda$ in the one-scale case). By \eqref{Tdecomp}, the tensor
$T^\mu{}_\nu$ has the eigenvalues $b\pm a$, that is $p\equiv a+b=\ell$ and $-\mathcal E$ with
\begin{equation}
\mathcal E\equiv 2\tau\ell'-\ell ,\qquad p\equiv\ell ,
\end{equation}
the energy density and the pressure. We use
$\kappa\equiv-b/a=r$, so that
\begin{equation}
\kappa=\frac{\mathcal E-p}{\mathcal E+p},\qquad \frac p{\mathcal E}=\frac{1-\kappa}{1+\kappa}.
\label{kappa}
\end{equation}
Under \eqref{causal} and $\ell(0)=0$, convexity gives $0=\ell(0)\ge\ell-\tau\ell'$, i.e.\ $b\le0$, and
$\ell>0$ for $\tau>0$. Hence
$\mathcal E\ge p>0$, $T^\mu{}_\mu\le0$ and $\kappa\in[0,1)$. The dominant energy condition
$\mathcal E\ge|p|$ is equivalent to $b\le0$ and so holds automatically in the causal sector.
The $T\bar T$ operator itself has the simple form
\begin{equation}
-\tfrac12\det T^\mu{}_\nu=\tfrac14(X-Y)=\tfrac12(a^2-b^2)=\tfrac12\,\mathcal E\,p .
\label{TTbarop}
\end{equation}

\subsection{Russo--Townsend auxiliary field approach}\label{sec:RT}

The Courant--Hilbert function $\ell(\tau)$ has an equivalent description in terms of a scalar
auxiliary field. Following Russo-Townsend~\cite{Russo:2024llm}, whose construction was carried
over to two dimensions in~\cite{Babaei-Aghbolagh:2025uoz,Baglioni:2025tsc} (see also~\cite{Gaillard:1997rt,Ivanov:2003uj} for Legendre transforms in self-dual electrodynamics), perform the Legendre transform
of $\ell$,
\begin{equation}
\ell(\tau)=-\Omega(y)+y\,\Omega_y(y),\qquad \ell'=y,\qquad \tau=\Omega_y(y),
\label{legendre}
\end{equation}
where $\Omega(y)=y\tau-\ell(\tau)$ is a function of the single real variable $y$, defined whenever
$\ell''\neq0$. Inserting \eqref{legendre} in \eqref{CHsol} the Lagrangian becomes
\begin{equation}
\mathcal L(u,v;y)=y\,v-\frac{u}{y}-\Omega(y).
\label{LRT}
\end{equation}
The stationarity condition $\partial_y\mathcal L=0$ is exactly the constraint $\tau=v+u/\ell'^2$ in
\eqref{CHsol}, because $\Omega_y=\tau$ and $y=\ell'$: the variable $y$ is a scalar auxiliary field
with a self-interaction potential $\Omega(y)$, and $\mathcal L(u,v)$ is obtained by eliminating it. The
PCM is the degenerate point $y\equiv1$, and 2D ModMax is $y\equiv e^\gamma$. 

\paragraph{Stress tensor in terms of the potential.}
From \eqref{legendre}, $\tau\ell'=y\Omega_y$ and $\ell-\tau\ell'=-\Omega$. Hence
\begin{equation}
T^\mu{}_\mu=-2\,\Omega,\qquad
X=2\left[(y\Omega_y)^2+\Omega^2\right],\qquad
Y=4\Omega^2,
\label{XYOmega}
\end{equation}
so that the invariants can be read off directly from the potential. Moreover, since
$\mathcal L$ is stationary in $y$, $\partial_\alpha\mathcal L|_{u,v}=-\partial_\alpha\Omega|_{y}$, which
reproduces Lemma~1 without any reference to $\tau$.

\paragraph{One-scale potentials.}
For the one-scale family \eqref{oneScale}, the Legendre transform has the same structure,
\begin{equation}
\Omega(y;\lambda,\gamma)=\frac{1}{\lambda}\,\omega(\hat y),\qquad
\hat y\equiv e^{-\gamma}y=f'(\zeta),\qquad
\omega(\hat y)=\hat y\,\zeta-f(\zeta),\quad\zeta=\omega'(\hat y).
\label{omega}
\end{equation}
Therefore, with $\Omega\propto\lambda^{-1}$ and $\hat y=e^{-\gamma}y$,
\begin{equation}
\frac{\partial\mathcal L}{\partial\lambda}=-\partial_\lambda\Omega\big|_y=\frac{\Omega}{\lambda},
\qquad
\frac{\partial\mathcal L}{\partial\gamma}=-\partial_\gamma\Omega\big|_y=y\,\Omega_y ,
\label{flowsRT}
\end{equation}
and the signed ratio \eqref{rrho} becomes
\begin{equation}
r=\frac{\Omega}{y\,\Omega_y}=\frac{\omega(\hat y)}{\hat y\,\omega'(\hat y)},\qquad
F=\frac{\Omega}{2\lambda\left[(y\Omega_y)^2+\Omega^2\right]}.
\label{rOmega}
\end{equation}
The flow function $\mathcal F(\rho)$ is thus obtained by eliminating $\hat y$ between the two
expressions in \eqref{rOmega}: the problem of finding the exact irrelevant operator
reduces to inverting one explicit function of one variable built out of the potential. The potentials
of the examples treated below are collected in table~\ref{tab:potentials}.

\paragraph{Weak-field expansion.}
Around the PCM point $\hat y=1$ the potential starts at second order. We write $W(\phi)\equiv\Omega(y)$ with $y \equiv e^{\phi}$,
\begin{equation}
W(\phi)=\Omega=\frac{1}{\lambda}\left(\omega_2\phi^2+\omega_3\phi^3+\omega_4\phi^4+\omega_5\phi^5+\cdots\right),
\label{WW2}
\end{equation}
with no linear term since $\Omega_y=\tau$ vanishes at the vacuum. The flow coefficients of \eqref{C012}
\begin{equation}
\begin{aligned}
C_0=&\frac{1}{8\,\omega_2},\qquad
C_1=-\frac{\omega_3}{2\sqrt2\,\omega_2^{2}},\qquad
C_2=- \frac{\omega_2^2 - 11 \omega_3^2}{16 \omega_2^3} -  \frac{3 \omega_4}{4 \omega_2^2},\\
C_3=&\frac{3 \omega_3 \omega_4}{\sqrt{2} \omega_2^3} -  \frac{10 \omega_3^3 -  \omega_2^2 (\omega_3 - 16 \omega_5)}{8 \sqrt{2} \omega_2^4}
\label{C01omega}
\end{aligned}
\end{equation}
and the coefficients are related to
those of $\ell$ by the Legendre transform become,
$\omega_2=\dfrac{1}{4a_1}$, $\omega_3=\dfrac{2a_1^2-a_2}{8a_1^3}$, $\omega_4=\frac{28 a_1^4 - 36 a_1^2 a_2 + 27 a_2^2}{192 a_1^5} -  \frac{a_3}{16 a_1^4}$, and $\omega_5=\frac{( 3 a_2-2 a_1^2) a_3}{16 a_1^6} + \frac{8 a_1^6 - 20 a_1^4 a_2 - 27 a_2^3 + 4 a_1^2 (9 a_2^2 -  a_4)}{128 a_1^7}$.
In the notation of the Introduction these are the relations between the coefficients $c_i$ of
$\mathcal F$ and the coefficients $b_i$ of the potential. The odd part of the flow vanishes at
leading order iff $\omega_3=-\omega_2$: for instance the Born--Infeld potential has
$(\omega_2,\omega_3)=(\tfrac12,-\tfrac12)$, whereas the logarithmic one has $(\tfrac12,-\tfrac13)$ and the
no-$\tau$-maximum one $(1,\tfrac13)$.

\begin{table}[t]
\centering\small
\begin{tabular}{@{}lll@{}}
\toprule
theory & $\omega(\hat y)$ & $\ell(\tau)=\lambda^{-1}f(\zeta)$\\
\midrule
GBI & $\tfrac12\left(\hat y+\hat y^{-1}-2\right)$ & $1-\sqrt{1-2\zeta}$\\
$q$-deformed & $q\hat y-(q-1)\hat y^{\,q/(q-1)}-1$ & $1-(1-\zeta/q)^q$\\
no-$\tau$-max. & $\tfrac13(\hat y-1)^2(\hat y+2)$ & $\tfrac23\left[(1+\zeta)^{3/2}-1\right]$\\
logarithmic & $\hat y-1-\ln\hat y$ & $-\ln(1-\zeta)$\\
\bottomrule
\end{tabular}
\caption{Russo--Townsend potentials $\Omega=\omega(e^{-\gamma}y)/\lambda$ of the examples of
sections~\ref{sec:GBI}--\ref{sec:log}. The no-$\tau$-maximum theory is the member $q=3/2$ of the
$q$-family with $\lambda'=-\tfrac32\lambda$; the GBI theory is $q=\tfrac12$ and the logarithmic one the limit
$q\to0$ with $\lambda=q\mu$ fixed.}
\label{tab:potentials}
\end{table}

\subsection{Relation to the auxiliary field sigma model and integrability}\label{sec:AFSM}

The potential $\Omega$ is the same object as the interaction function of the auxiliary field sigma
model (AFSM) of Ferko and Smith~\cite{Ferko:2024ali}.
In its $\nu$-frame the AFSM is $\mathcal L=\tfrac12\tr(j_+j_-)+\tr(v_+v_-)+\tr(j_+v_-+j_-v_+)+E(\nu)$,
$\nu=\sqrt{\tr(v_+^2)\tr(v_-^2)}$, whose conserved current is $\mathfrak J_\pm=-(j_\pm+2v_\pm)$ and whose Lax
connection is $\mathfrak L_\pm=(j_\pm\pm z\mathfrak J_\pm)/(1-z^2)$. The Legendre transform
$H(\mu)=E-\nu E_\nu$, $\mu=E_\nu$ (the $\mu$-frame) is identified with the RT potential
by~\cite{Baglioni:2025tsc}
\begin{equation}
y=\frac{1+\mu}{1-\mu},\qquad H(\mu)=-\Omega(y),
\qquad
\ell'=\frac{1+E_\nu}{1-E_\nu},\quad \tau=\frac\nu2(1-E_\nu)^2 .
\label{muframe}
\end{equation}
As a check, 2D ModMax $\ell=e^\gamma\tau$ has $\ell'=e^\gamma$ and therefore $E_\nu=\tanh(\gamma/2)$,
i.e.\ $E=\tanh(\gamma/2)\,\nu$, which is the known solution of the root-$T\bar T$
flow~\cite{Bielli:2024ach}. For the Born--Infeld potential of table~\ref{tab:potentials} one finds
from \eqref{muframe} that $H(\mu)=-\Omega_{\rm GBI}$ coincides with the solution of the mixed
$T\bar T$ and root-$T\bar T$ flows of~\cite{Baglioni:2025tsc}, provided one takes into account that their
$T\bar T$ coupling is normalised by $\partial_{\lambda}\mathcal L=T_{\mu\nu}T^{\mu\nu}-(T^\mu{}_\mu)^2=X-Y$,
i.e.\ $\lambda_{\rm here}=4\lambda_{\rm there}$ (see \eqref{GBIflow}).

Because the Lax connection and the Maillet bracket of the AFSM do not depend on $E$ \cite{Ferko:2024ali,Bielli:2024ach}
(the twist function $\varphi(z)=(z^2-1)/z^2$ is that of the PCM), every theory in this section is classically
integrable, in the strong sense of an infinite set of charges in involution. The Yangian
structure of these models has been studied in~\cite{Bielli:2026ggs}. The flows
$\partial_\lambda\mathcal L=X\mathcal F(Y/X)$ and $\partial_\gamma\mathcal L=\tau\ell'$ are therefore
integrability-preserving deformations of the PCM, and $X\mathcal F(Y/X)$ is the exact form of
the operator $\mathcal O_\lambda$ of the Introduction. We now turn to the examples.

\section{Generalised Born--Infeld sigma model}\label{sec:GBI}
\subsection{Exact flow}

The simplest non-trivial one-scale solution is the generalised Born--Infeld (GBI) model,
\begin{equation}
\ell_{\rm GBI}(\tau;\lambda,\gamma)=\frac1\lambda\Big(1-\sqrt{1-2e^\gamma\lambda\tau}\Big),
\qquad f(\zeta)=1-\sqrt{1-2\zeta}.
\label{ellGBI}
\end{equation}
It is the two-dimensional Born--Infeld-like deformation of the PCM (the auxiliary-field model of
\cite{Ferko:2024ali}, see table~\ref{tab:potentials}), and $\gamma$ is the
ModMax-like parameter. With
\begin{equation}
s\equiv\sqrt{1-2\zeta},\qquad \zeta=\tfrac12(1-s^2),\qquad f=1-s,\qquad f'=\frac1s,
\end{equation}
the functions \eqref{ab} are
\begin{equation}
a(s)=\frac{1-s^2}{2s},\qquad b(s)=-\frac{(1-s)^2}{2s},
\end{equation}
and \eqref{Fzeta} gives the exact parametric flow function
\begin{equation}
F_{\rm GBI}(s)=\frac{s}{2(1+s^2)} .
\end{equation}
The auxiliary variable is eliminated algebraically. The signed ratio is
\begin{equation}
r(s)=-\frac ba=\frac{1-s}{1+s},\qquad s(r)=\frac{1-r}{1+r},
\end{equation}
so that $F_{\rm GBI}(r)=\dfrac{1-r^2}{4(1+r^2)}$, and since $r^2/(1+r^2)=\rho^2/2$ by \eqref{rrho},
\begin{equation}
\mathcal F_{\rm GBI}(\rho)=\frac{1-\rho^2}{4}.
\end{equation}
The autonomous flow \eqref{autonomous} is therefore a \emph{polynomial} deformation, quadratic in the
stress tensor,
\begin{equation}
\ \frac{\partial\mathcal L_{\rm GBI}}{\partial\lambda}
=\frac14\,T_{\mu\nu}T^{\mu\nu}-\frac14\big(T^\mu{}_\mu\big)^2=-\frac12\det T^\mu{}_\nu\ 
\label{GBIflow}
\end{equation}
(recall $X=T_{\mu\nu}T^{\mu\nu}$, $Y=(T^\mu{}_\mu)^2$). This is the $T\bar T$ operator of
Zamolodchikov, Smirnov and Cavagli\`a et al.~\cite{Zamolodchikov:2004ce,Smirnov:2016lqw,Cavaglia:2016oda}, in the form that generates generalised Born--Infeld models~\cite{Conti:2018jho,Born:1934gh},
and the combination of \eqref{GBIflow} with $\partial_\gamma\mathcal L=\tfrac12\sqrt{2X-Y}$ reproduces the
mixed $T\bar T$/root-$T\bar T$ flow of the auxiliary field sigma models~\cite{Ferko:2024ali,Bielli:2024ach,Baglioni:2025tsc}.
Thus, although $\mathcal L_{\rm GBI}$ is a non-polynomial function of the currents, its exact flow is a
quadratic operator, and all half-integer powers in \eqref{Cm} vanish identically.

\subsection{Physical domain and interpretation}

For $\lambda>0$ and $\gamma\ge0$ reality of \eqref{ellGBI} requires
\begin{equation}
0\le\tau\le\frac1{2e^\gamma\lambda},\qquad 0\le\zeta\le\tfrac12,\qquad 0\le s\le1,
\qquad 0\le r\le1,\qquad 0\le\rho\le1 .
\end{equation}
The expansion of $\mathcal F_{\rm GBI}$ is exact at second order,
\begin{equation}
C_0=\tfrac14,\qquad C_1=0,\qquad C_2=-\tfrac14,\qquad C_{m\ge3}=0,
\end{equation}
in agreement with \eqref{C012} for $f=1-\sqrt{1-2\zeta}$ ($a_1=\tfrac12$, $a_2=2a_1^2$).
In terms of the eigenvalues of section~\ref{sec:stress}, \eqref{GBIflow} reads
\begin{equation}
\frac{\partial\mathcal L_{\rm GBI}}{\partial\lambda}=\tfrac12(a^2-b^2)=\tfrac12\,\mathcal E\,p
=\tau\ell\ell'-\tfrac12\ell^2 ,
\label{GBIphys}
\end{equation}
a polynomial in $(\ell,\tau\ell')$ without any singularity on the causal window. The critical value
$\tau_{\max}=1/(2e^\gamma\lambda)$ is the analogue of the Born--Infeld bound: at $\tau\to\tau_{\max}$,
$\ell'\to\infty$ and $\mathcal E\to\infty$ while $p\to1/\lambda$. The GBI model is the member $q=\tfrac12$
of the family of the next section and the only member whose flow is polynomial in the stress tensor.

\section{The $q$-deformed sigma models}\label{sec:q}
\subsection{Exact non-perturbative $T\bar T$-like flow}

The $q$-deformed family is the two-dimensional version of the $q$-deformed ModMax theories of Russo-Townsend~\cite{Russo:2024ptw}. Consider the one-parameter family
\begin{equation}
\ell_q(\tau;\lambda,\gamma)=\frac1\lambda\Big[1-\Big(1-\frac{e^\gamma\lambda\tau}{q}\Big)^{q}\Big],
\qquad s\equiv1-\frac\zeta q,\qquad f_q(s)=1-s^q ,
\label{ellq}
\end{equation}
which reduces to $\ell=e^\gamma\tau$ (2D ModMax) at $q=1$. One has
$\ell'=e^\gamma s^{q-1}$ and $\ell''=e^{2\gamma}\lambda\,\frac{1-q}{q}\,s^{q-2}$, so that the causality
conditions \eqref{causal} with $\gamma\ge0$, $q>0$ select the branches
\begin{center}
\begin{tabular}{c|c|c|c}
$q$ & $\lambda$ & $s$ & $\tau$\\\hline
$0<q<1$ & $\lambda>0$ & $0<s\le1$ & $0\le\tau<qe^{-\gamma}/\lambda$\\
$q>1$ & $\lambda<0$ & $1\le s<\infty$ & $0\le\tau<\infty$
\end{tabular}
\end{center}
In particular the theories with $q>1$ are causal only for negative $\lambda$. Near the PCM point
\begin{equation}
\ell_q=e^\gamma\tau+c_q\lambda e^{2\gamma}\tau^2+O(\tau^3),\qquad c_q=\frac{1-q}{2q},
\end{equation}
i.e.\ $a_1=c_q$, so that $c_q\lambda\ge0$ is the sign of the quartic term
$\mathcal L=v-u+c_q\lambda(u+v)^2+\dots$ at $\gamma=0$.

The functions \eqref{ab} are $a_q=q(1-s)s^{q-1}$, $b_q=1-qs^{q-1}+(q-1)s^q$, and therefore
\begin{equation}
r_q(s)=\frac{q+(1-q)s-s^{1-q}}{q(1-s)} .
\label{rq}
\end{equation}
At the vacuum $s=1$ this is $0/0$, but the singularity is removable and $r_q(1)=0$. On each physical
branch $r_q$ is strictly monotonic, hence invertible, and the inverse is fixed by
\begin{equation}
s_q^{\,1-q}=q+(1-q)s_q-q(1-s_q)\,r ,
\label{sqdef}
\end{equation}
where $s=1$ is a spurious root for $r\ne0$ and the physical one is selected by $0<s_q<1$ ($q<1$) or
$s_q>1$ ($q>1$). The ranges are $0<r<1$ for $q<1$ and $0<r<(q-1)/q$ for $q>1$. The flow function is
\begin{equation}
F_q(r)=\frac{r\,s_q^{1-q}}{2q(1-s_q)(1+r^2)},\qquad
\mathcal F_q(\rho)=\frac{\rho\sqrt{2-\rho^2}}{4q}\,\frac{s_q^{1-q}}{1-s_q},
\label{Fq}
\end{equation}
with $r=\rho/\sqrt{2-\rho^2}$, and
\begin{equation}
\frac{\partial\mathcal L_q}{\partial\lambda}=X\,\mathcal F_q\!\Big(\sqrt{Y/X}\Big),
\end{equation}
on the domains
\begin{equation}
0<\rho<1\ \ (0<q<1),\qquad
0<\rho<\frac{\sqrt2\,(q-1)}{\sqrt{q^2+(q-1)^2}}\ \ (q>1).
\end{equation}
No explicit dependence on $\lambda,\gamma,\tau,u,v$ survives: the flow is exact, non-perturbative and
autonomous. Its weak-field expansion $\mathcal F_q=\sum_m C_m(q)\rho^m$ has the coefficients
\begin{equation}
\begin{aligned}
C_0&=\frac{1-q}{4q},\qquad C_1=\frac{\sqrt2\,(2q-1)}{6q},\qquad
C_2=-\frac{(q+1)(8q-7)}{72\,q(q-1)},\\
C_3&=-\frac{\sqrt2\,(2q-1)(43q^2-88q+49)}{1080\,q(q-1)^2},\qquad
C_4=\frac{(2q-1)^2(q-2)(q+1)}{1620\,q(q-1)^3},\\
C_5&=-\frac{\sqrt2\,(2q-1)\big(2335q^4-10088q^3+16926q^2-12512q+3499\big)}{544320\,q(q-1)^4},\\
C_6&=\frac{(2q-1)^2(q-2)(q+1)(331q^2-541q+268)}{680400\,q(q-1)^5}.
\end{aligned}
\label{Cq}
\end{equation}
Compared with the four-dimensional analogue \cite{Pezhman:2026dsr,Pezhman:2026sby,Pezhman:2026gbv}, $C_m^{\rm 2D}=2^{1+m/2}C_m^{\rm 4D}$:
the dimensional reduction only rescales the invariants.\footnote{We have checked \eqref{Cq} against the
general formulas \eqref{C012} for $m\le2$ and, for $m\le6$, by exact rational series arithmetic and by
numerical comparison of $X\mathcal F_q$ with a finite-difference evaluation of $\partial_\lambda\mathcal L_q$.}
Only $q=\tfrac12$ and $q=1$ have $C_1=0$.

\subsection{The cases $q=\tfrac12,\ \tfrac34,\ \tfrac32$}

\paragraph{$q=\tfrac12$.} With $w=\sqrt s$, \eqref{rq} gives $r=(1-w)/(1+w)$, hence $s=\big(\tfrac{1-r}{1+r}\big)^2$ and
$\mathcal F_{1/2}=\tfrac14(1-\rho^2)$: this is the GBI flow \eqref{GBIflow}; all $C_{m}$ in \eqref{Cq} vanish except
$C_0=\tfrac14$, $C_2=-\tfrac14$.

\paragraph{$q=\tfrac34$.} With $w=s^{1/4}$, $r=\dfrac{3+w^4-4w}{3(1-w^4)}$, and the physical root of
\begin{equation}
(1+3r)(w^3+w^2+w)=3(1-r),\qquad w\in(0,1),
\end{equation}
gives $F_{3/4}(r)=\dfrac{2\,r\,w}{3(1-w^4)(1+r^2)}$ and
\begin{equation}
\mathcal F_{3/4}(\rho)=\frac1{12}+\frac{\sqrt2}{9}\rho-\frac{7}{54}\rho^2-\frac{23\sqrt2}{324}\rho^3
+\frac{7}{243}\rho^4-\frac{869\sqrt2}{23328}\rho^5+\frac{155}{2916}\rho^6+O(\rho^7),
\end{equation}
which is \eqref{Cq} at $q=\tfrac34$. The odd coefficients are non-zero: the flow contains half-integer
powers of $Y/X$.

\paragraph{$q=\tfrac32$.} This member is on the branch $q>1$, causal for $\lambda<0$. It coincides with the
theory of section~\ref{sec:NM} with $\lambda_q=-\tfrac32\lambda_{\rm NM}$, $\lambda_{\rm NM}>0$, whose flow is available in closed form.

\subsection{Lemma 2 and the an identity}

\emph{Lemma 2.} For \eqref{ellq} the non-trivial root of \eqref{sqdef} is exactly $s_q=1-\zeta/q$ and
\begin{equation}
\frac{\partial\mathcal L_q}{\partial\lambda}=\frac{\tau\ell'-\ell}{\lambda}=-\frac{T^\mu{}_\mu}{2\lambda}
=\frac1{\lambda^2}\Big[q(1-s)s^{q-1}+s^q-1\Big].
\label{Lemma2}
\end{equation}
\emph{Proof.} By \eqref{ellq}, $\ell=(1-s^q)/\lambda$ and $a=\tau\ell'=q(1-s)s^{q-1}/\lambda$; substituting
$s_q=s$ in \eqref{sqdef} gives an identity, because $r=(a-\ell)/a$. Then $X=2a^2(1+r^2)$ turns \eqref{Fq} into
$X F_q=a^2 r\,s^{1-q}/[q(1-s)]=a\,r/\lambda=(a-\ell)/\lambda$, which equals $\partial_\lambda\ell_q|_\tau$ by direct differentiation. $\square$

Hence \eqref{Lemma2} is the an identity for the dimensionful coupling; what depends on $q$ is
how $T^\mu{}_\mu/\lambda$ is written through the invariants of $T_{\mu\nu}$ alone, which requires solving
\eqref{sqdef}. The $q\to0$ limit with $\lambda=q\mu$ fixed produces the logarithmic model of
section~\ref{sec:log}, the limit $q\to\infty$ the exponential model ($\lambda<0$).

\subsection{Singular points}

All singular points lie on the one-dimensional window $\kappa=r\in[0,1)$.

\paragraph{(i) $r\to0$, the conformal point.} Here $\zeta\to0$, $T^\mu{}_\mu\to0$, $p\to\mathcal E$ and
$\partial_\lambda\mathcal L_q\simeq c_q\,e^{2\gamma}\tau^2$: this is the quartic seed coupling. In the variable $\tau$ the
operator is analytic; the odd powers of $\rho\propto|T^\mu{}_\mu|$ are an artefact of using $r\ge0$ as a
coordinate (on the causal domain $T^\mu{}_\mu\le0$ and $|T^\mu{}_\mu|$ is analytic).

\paragraph{(ii) $r\to1$ for $q<1$, the critical current.} Here $s\to0$, $\tau\to qe^{-\gamma}/\lambda$ and $\ell'\to\infty$.
Since $\tau=v+u/\ell'^2$, $\mathcal L$ is defined for all $u\ge0$ iff
$v<v_{\max}=q\,e^{-\gamma}/\lambda$: the combination $v$ of the currents is bounded, the 2D analogue of the
Born--Infeld bound. At the bound $\mathcal E\to\infty$, $p\to1/\lambda$, $p/\mathcal E\to0^+$ and
$\partial_\lambda\mathcal L_q\simeq\mathcal E/(2\lambda)$. In terms of $\epsilon=1-r$, $s_q\simeq(q\epsilon)^{1/(1-q)}$ and the
endpoint is a branch point $\propto s^q$ of $\ell_q$ (a square root at $q=\tfrac12$). Beyond it $\mathcal L$ is complex,
which marks the limit of validity of the effective description.

\paragraph{(iii) $r\to r_c=(q-1)/q$ for $q>1$.} Here $\lambda<0$, $s\to\infty$ as $\tau\to\infty$; there is no critical current,
$\ell_q\propto\tau^q$, $p/\mathcal E\to1/(2q-1)$, and $\partial_\lambda\mathcal L_q\simeq-(q-1)s^q/\lambda^2$. The absence of a real
root for $r>r_c$ only says that the family does not realise stress tensors with $p/\mathcal E<1/(2q-1)$.

\paragraph{(iv) Sign of the flow.} Since $T^\mu{}_\mu\le0$, $\partial_\lambda\mathcal L_q$ has the sign of $\lambda$.

\begin{table*}[t]
\centering\small
\begin{tabular}{@{}llllp{4.4cm}@{}}
\toprule
Point & Fields & Stress tensor & $\partial_\lambda\mathcal L_q$ & Interpretation\\
\midrule
$r\to0$ & $\tau\to0$ & $T^\mu{}_\mu\to0,\ p\to\mathcal E$ & $c_qe^{2\gamma}\tau^2$ & conformal point, quartic coupling $c_q\lambda\ge0$\\
$r\to1$, $q<1$ & $\tau\to\frac{q}{\lambda e^\gamma}$ & $\mathcal E\to\infty,\ p\to1/\lambda$ & $\simeq\mathcal E/2\lambda$ & critical current, branch point $s^q$\\
$r\to\frac{q-1}q$, $q>1$ & $\tau\to\infty$ & $p/\mathcal E\to\frac1{2q-1}$ & $\simeq-\frac{(q-1)s^q}{\lambda^2}$ & strong-field regime, no critical current\\
$q=\frac12$ & --- & $\tfrac12\mathcal E p$ & $=\partial_\lambda\mathcal L_{\rm GBI}$ & only polynomial member\\
$q=1$ & $\ell=e^\gamma\tau$ & $r\equiv0$ & $0$ & 2D ModMax, $\lambda$ drops out\\
\bottomrule
\end{tabular}
\caption{Special points of the $q$-deformed flow.}
\label{tab:qsing}
\end{table*}

\section{The no-$\tau$-maximum sigma model}\label{sec:NM}
\subsection{Courant--Hilbert function and exact $T\bar T$-like flow}

The model
\begin{equation}
\ell_{\rm NM}=\frac2{3\lambda}\Big[(1+\zeta)^{3/2}-1\Big],\qquad s=\sqrt{1+\zeta}\ge1,\qquad f=\tfrac23(s^3-1),\ f'=s,
\end{equation}
has $\ell'=e^\gamma s\ge1$ and $\ell''=e^{2\gamma}\lambda/(2s)>0$ for all $\tau\ge0$, $\lambda>0$, $\gamma\ge0$, so it is causal with no
critical current. It is the member $q=\tfrac32$ of \eqref{ellq} with $\lambda_q=-\tfrac32\lambda$. The functions \eqref{ab} are
\begin{equation}
a=s(s^2-1),\qquad b=-\tfrac13(s-1)^2(s+2),
\end{equation}
so that
\begin{equation}
r(s)=\frac13-\frac2{3s(s+1)},\qquad \frac{dr}{ds}=\frac{2(2s+1)}{3s^2(s+1)^2}>0,\qquad 0\le r<\tfrac13 ,
\end{equation}
where $r(1)=0$ by continuity. Since $r$ is monotonic the inverse is explicit,
$s(r)=\tfrac12\Big[\sqrt{3(3-r)/(1-3r)}-1\Big]$, and
\begin{equation}
F_{\rm NM}(r)=\frac{(1-3r)^2}{48(1+r^2)}\Bigg[3+\sqrt{\frac{3(3-r)}{1-3r}}\Bigg],\qquad
r=\frac{\rho}{\sqrt{2-\rho^2}},\quad 0\le\rho<\frac1{\sqrt5},
\end{equation}
\begin{equation}
\frac{\partial\mathcal L_{\rm NM}}{\partial\lambda}=X\,\mathcal F_{\rm NM}(\rho),\qquad
\mathcal F_{\rm NM}=\frac18-\frac{\sqrt2}3\rho+\frac{25}{72}\rho^2+\frac{11\sqrt2}{108}\rho^3+\frac2{81}\rho^4
+\frac{269\sqrt2}{7776}\rho^5+\frac{23}{486}\rho^6+\dots
\end{equation}
(equal to $-\tfrac32$ times the $q=\tfrac32$ series of \eqref{Cq}, because $\partial_{\lambda_{\rm NM}}=-\tfrac32\partial_{\lambda_q}$). In stress-tensor variables
\begin{equation}
\frac{\partial\mathcal L_{\rm NM}}{\partial\lambda}=\frac{(s-1)^2(s+2)}{3\lambda^2},\qquad
\frac p{\mathcal E}=\frac{1-r}{1+r}=\frac{s^2+s+1}{2s^2+2s-1}\in\big(\tfrac12,1\big].
\end{equation}

\subsection{Singular points}
\begin{itemize}
\item $r=0$: $r\simeq(s-1)/2$ is analytic and $\partial_\lambda\mathcal L_{\rm NM}\to\tfrac14e^{2\gamma}\tau^2$, i.e.\ the quartic coefficient $\lambda/4>0$ ($a_1=\tfrac14$, $C_0=\tfrac18$).
The odd powers of $\rho$ are analytic on the causal domain where $T^\mu{}_\mu\le0$.
\item $r\to\tfrac13$ ($\rho\to1/\sqrt5$): $\mathcal F_{\rm NM}\simeq\frac{3\sqrt2}{80}(1-3r)^{3/2}\simeq\frac{3}{20s^3}$, a branch point of the
invariants corresponding to $s,\tau\to\infty$, not to a finite field strength; it says only that no stress tensor with $p/\mathcal E<\tfrac12$ is realised
($\ell\propto\tau^{3/2}$ there).
\item $1+r^2$ never vanishes for real $r$: there are no poles.
\end{itemize}

\section{The logarithmic sigma model}\label{sec:log}
\subsection{Exact non-perturbative $T\bar T$-like flow}

The logarithmic solution, the analogue of logarithmic electrodynamics~\cite{Bialynicki-Birula:1984daz,Russo:2024kto}, is
\begin{equation}
\ell_{\log}=-\frac1\lambda\ln(1-\zeta),\qquad f=-\ln(1-\zeta),\qquad f'=\frac1{1-\zeta},
\end{equation}
which is the limit $q\to0$ of \eqref{ellq} with $\lambda=q\mu$ held fixed ($\mu\to\lambda$). Its causal
domain is $\lambda>0$, $\gamma\ge0$, $0\le\zeta<1$. With $s=1-\zeta$,
\begin{equation}
a=\frac{1-s}{s},\qquad b=1-\frac1s-\ln s,\qquad r(s)=1+\frac{s\ln s}{1-s}.
\end{equation}
The physical inverse is given by the lower real branch $W_{-1}$ of the Lambert function~\cite{Corless:1996},
\begin{equation}
s(r)=\frac{r-1}{W_{-1}\big((r-1)e^{r-1}\big)},\qquad 0<s<1,
\end{equation}
(the principal branch gives the trivial root $s=1$, physical only at the vacuum, where the branches meet). Then
\begin{equation}
F_{\log}(r)=\frac{r\,s(r)}{2[1-s(r)](1+r^2)},\qquad \frac{\partial\mathcal L_{\log}}{\partial\lambda}=X\,\mathcal F_{\log}(\rho),
\end{equation}
\begin{equation}
\mathcal F_{\log}=\frac14-\frac{\sqrt2}{6}\rho-\frac7{72}\rho^2+\frac{49\sqrt2}{1080}\rho^3+\frac1{810}\rho^4
+\frac{3499\sqrt2}{544320}\rho^5+\frac{67}{85050}\rho^6+\dots,
\end{equation}
the $q\to0$ limit of \eqref{Cq} after multiplication by $q$. The physical intervals are $0<s<1$, $0<\zeta<1$,
$0<r<1$ and $0<\rho<1$.

\subsection{Compact form and analytic structure}

Put $\varsigma\equiv r-1$ and $W\equiv W_{-1}(\varsigma e^{\varsigma})$, so that $1-s=(W-\varsigma)/W$. Using $X/(1+r^2)=2a^2$,
\begin{equation}
 \frac{\partial\mathcal L_{\log}}{\partial\lambda}=(\tau\ell')^2\,\frac{r(r-1)}{D(r)},\qquad
D(r)=W_{-1}\big((r-1)e^{r-1}\big)-(r-1),
\label{logcompact}
\end{equation}
and since $r-1=-\ell/(\tau\ell')$, explicitly
\begin{equation}
\frac{\partial\mathcal L_{\log}}{\partial\lambda}=\frac{\ell\,(\ell-\tau\ell')}
{W_{-1}\!\big(-\tfrac{\ell}{\tau\ell'}e^{-\ell/\tau\ell'}\big)+\tfrac{\ell}{\tau\ell'}}.
\end{equation}
On the dilatation side one has the elementary expression
\begin{equation}
\frac{\partial\mathcal L_{\log}}{\partial\lambda}=\frac{\tau\ell'-\ell}{\lambda}=\frac1{\lambda^2}\Big[\frac\zeta{1-\zeta}+\ln(1-\zeta)\Big],
\end{equation}
so the Lambert function arises only from expressing $\lambda$ through the invariants of $T_{\mu\nu}$. (Numerical check at
$\gamma=0$, $\lambda=1$, $\tau=\tfrac12$: both expressions give $0.3069$.)

\subsection{Singular points}

All singularities of \eqref{logcompact} come from the Lambert function: $W_{-1}(\varsigma e^\varsigma)$ is real only for $\varsigma<0$.
\begin{itemize}
\item $r<0$: here $W_{-1}(\varsigma e^\varsigma)=\varsigma$, $D\equiv0$, and the literal formula is undefined. This is the region $b>0$, excluded in the causal sector by convexity.
\item $r=0$: the argument of $W$ is $-1/e$, where the branches meet. Setting $r=\epsilon>0$, $W_{-1}=-1-\epsilon+O(\epsilon^3)$ and $D=-2\epsilon+O(\epsilon^3)$, so the $0/0$ cancels:
$\partial_\lambda\mathcal L_{\log}\to\tfrac12(\tau\ell')^2$, equal to the quartic seed term ($a_1=\tfrac12$, $C_0=\tfrac14$). This is a coordinate effect; in $\zeta$ the operator is analytic.
\item $r\to1^-$: $W_{-1}(x)\simeq\ln(-x)-\ln(-\ln(-x))\to-\infty$ and
$\partial_\lambda\mathcal L_{\log}\simeq(\tau\ell')^2(1-r)/|\ln(1-r)|\to0^+$: a logarithmic branch point, continuous with vanishing first derivative. It is the critical-current point
$v\to v_{\max}=e^{-\gamma}/\lambda$ ($\zeta\to1$, $\mathcal E\simeq2/[\lambda(1-\zeta)]\to\infty$, $p=\ell\to\infty$ only logarithmically, $p/\mathcal E\to0^+$). For $r>1$, $\ell<0$ and the argument of $W_{-1}$ is positive: not real.
\item $0<r<1$: $W_{-1}<-1<\varsigma$, so $D<0$ and the flow is smooth and pole-free; $1+r^2$ has no real zeros.
\end{itemize}

\begin{table*}[t]
\centering\small
\begin{tabular}{@{}llllp{4.3cm}@{}}
\toprule
Point & Fields & Stress tensor & $\partial_\lambda\mathcal L_{\log}$ & Interpretation\\
\midrule
$r=0$ & $\zeta\to0$, $r\simeq\zeta/2$ & $T^\mu{}_\mu\to0,\ p\to\mathcal E$ & $\tfrac12e^{2\gamma}\tau^2$ & conformal point, removable\\
$0<r<1$ & $0<\zeta<1$ & $\mathcal E\ge p>0$ & smooth, $D<0$ & causal window\\
$r\to1^-$ & $v\to e^{-\gamma}/\lambda$ & $\mathcal E\to\infty$, $p\sim\ln\frac1{1-\zeta}$ & $\sim\epsilon/|\ln\epsilon|$ (normalised) & critical current, log branch point\\
$r<0$ & $\lambda<0$ ($W_0$) & --- & $D\equiv0$ & non-causal\\
$r>1$ & $\ell<0$ & --- & not real & outside the theory\\
\bottomrule
\end{tabular}
\caption{Special points of the logarithmic flow ($\epsilon=1-r$).}
\label{tab:logsing}
\end{table*}

\section{Flows in arbitrary dimension: from $d=2$ to $d=4$}\label{sec:anyd}

The two-dimensional flow \eqref{Fzeta} and its four-dimensional counterpart for self-dual nonlinear
electrodynamics~\cite{Babaei-Aghbolagh:2025cni,Pezhman:2026dsr,Pezhman:2026sby,Pezhman:2026gbv} differ only by numerical factors; this is the dimensional counterpart of the relation between 2D integrable models and 4D duality-invariant electrodynamics found in~\cite{Babaei-Aghbolagh:2025uoz,Baglioni:2025tsc}. In this section we show that they are the
two members $d=2,4$ of a single flow equation written for a spacetime of dimension $d$, and we
explain the relation $C_m^{\rm 2D}=2^{1+m/2}C_m^{\rm 4D}$ found in section~\ref{sec:q}.

\subsection{Stress tensor and invariants in $d$ dimensions}

Consider a theory in $d$ dimensions whose Courant--Hilbert function depends on one dimensionful coupling,
\begin{equation}
\ell=\lambda^{-1}f(\zeta),\qquad \zeta=e^{\gamma}\lambda\tau,\qquad [\lambda]=-d,\quad [\mathcal L]=[\tau]=d ,
\end{equation}
and whose stress tensor has the structure
\begin{equation}
T_{\mu\nu}=\tau\ell'\,S_{\mu\nu}+(\ell-\tau\ell')\,g_{\mu\nu},\qquad S^\mu{}_\mu=0,\qquad S_{\mu\nu}S^{\mu\nu}=d .
\label{Tgend}
\end{equation}
Here $S_{\mu\nu}$ is a traceless tensor built from the fields, and $\ell'=\partial_\tau\ell$. It has the eigenvalues $\pm1$ with equal degeneracy~$d/2$. For $d=2$ the tensor $S_{\mu\nu}$ is the traceless part of the sigma-model stress tensor, so that $T^\mu{}_\nu$ has the eigenvalues $p=\ell$ and $-\mathcal E=-(2\tau\ell'-\ell)$ of section~\ref{sec:stress}. For $d=4$ it is $S_{\mu\nu}=T^{\rm Max}_{\mu\nu}/(U+V)$, and \eqref{Tgend} is the decomposition of self-dual nonlinear electrodynamics. We take \eqref{Tgend} as the defining property of the class considered here. We do not claim that it is realised in every dimension, but only that, whenever it is, the results below follow.

With $a=\zeta f'$ and $b=f-\zeta f'$ we have $\tau\ell'=a/\lambda$ and $\ell-\tau\ell'=b/\lambda$, and \eqref{Tgend} gives
\begin{equation}
T^\mu{}_\mu=\frac{d\,b}{\lambda},\qquad
X\equiv T_{\mu\nu}T^{\mu\nu}=\frac{d\,(a^2+b^2)}{\lambda^2},\qquad
Y\equiv\big(T^\mu{}_\mu\big)^2=\frac{d^2\,b^2}{\lambda^2},
\label{XYd}
\end{equation}
so that
\begin{equation}
\frac{Y}{X}=\frac{d\,b^2}{a^2+b^2}\in[0,d],\qquad
\tau\ell'=\frac{a}{\lambda}=\frac{\sqrt{dX-Y}}{d}.
\label{ratiod}
\end{equation}
For $d=2$ these are $X=2(a^2+b^2)/\lambda^2$ and $Y=4b^2/\lambda^2$. For $d=4$ they are
$X=4(a^2+b^2)/\lambda^2$ and $Y=16b^2/\lambda^2$.

\subsection{The flow equation for arbitrary $d$}

Since $\partial_\lambda\mathcal L=-b/\lambda^2$ and $\partial_\gamma\mathcal L=a/\lambda$ (Lemma~1 holds in any dimension, because it only uses the Courant--Hilbert form), the single trace flow reads $\partial_\lambda\mathcal L=-T^\mu{}_\mu/(d\lambda)$, and dividing by $X$ from \eqref{XYd} gives
\begin{equation}
\frac{\partial\mathcal L}{\partial\lambda}=X\,F_d(\zeta),\qquad
F_d(\zeta)=-\frac{b(\zeta)}{d\,\big[a(\zeta)^2+b(\zeta)^2\big]}\ ,
\qquad
 \frac{\partial\mathcal L}{\partial\gamma}=\frac{\sqrt{dX-Y}}{d}\ .
\label{flowd}
\end{equation}
For $d=2$ the first equation is \eqref{Fzeta}, and the second is \eqref{gammaflow}. For $d=4$ they are
the equations $F=-b/\{4[a^2+b^2]\}$ and $\partial_\gamma\mathcal L=\tfrac14\sqrt{4X-Y}$ of~\cite{Pezhman:2026dsr,Pezhman:2026sby,Pezhman:2026gbv}; in the 4D paper the first is
eq.~(25) and in the present paper it is eq.~(22). The dimensions are consistent for every $d$: $[\partial_\lambda\mathcal L]=2d=[X]=[Y]$.

As in section~\ref{sec:CH}, introduce the signed ratio $r=-b/a=-T^\mu{}_\mu/\sqrt{dX-Y}$ and $\rho=\sqrt{Y/X}$. Then
\begin{equation}
\rho^2=\frac{d\,r^2}{1+r^2},\qquad r=-\operatorname{sgn}(T^\mu{}_\mu)\frac{\rho}{\sqrt{d-\rho^2}},\qquad
F_d=\frac{r}{d\,a\,(1+r^2)} .
\label{rrhod}
\end{equation}
At the conformal point $F_d(0)=f''(0)/(2d)$, which equals $f''(0)/4$ for $d=2$ and $f''(0)/8$ for $d=4$. Whenever $\zeta\mapsto\rho^2$ is one-to-one on the physical branch, the flow is autonomous,
\begin{equation}
\frac{\partial\mathcal L}{\partial\lambda}=X\,\mathcal F_d(\rho)=\sum_{m\ge0}C_m^{(d)}\,Y^{m/2}X^{1-m/2}.
\label{autod}
\end{equation}

\subsection{A dimension-independent function}

Define the rescaled variable $\sigma\equiv\rho/\sqrt d=\sqrt{Y/(dX)}\in[0,1]$. Then $r=\sigma/\sqrt{1-\sigma^2}$ in every dimension, so by \eqref{rrhod}
\begin{equation}
 \frac{\partial\mathcal L}{\partial\lambda}=\frac{X}{d}\,\mathcal G(\sigma),\qquad
\mathcal G=\frac{r}{a(1+r^2)}\ ,
\label{Guniv}
\end{equation}
where $\mathcal G$ is determined by the function $f$ alone: given $f$, the function $\mathcal G(\sigma)$ is the same for every $d$. The dimension enters only through the prefactor $X/d$ and through the relation $\sigma^2=Y/(dX)$. Expanding $\mathcal G=\sum_m g_m\sigma^m$ and using $\rho^m=d^{m/2}\sigma^m$,
\begin{equation}
C_m^{(d)}=d^{-1-m/2}\,g_m,\qquad\text{so}\qquad
\frac{C_m^{(d_1)}}{C_m^{(d_2)}}=\Big(\frac{d_2}{d_1}\Big)^{1+m/2}.
\label{Cd}
\end{equation}
For $(d_1,d_2)=(2,4)$ this is the relation $C_m^{\rm 2D}=2^{1+m/2}C_m^{\rm 4D}$. With $f=\zeta+a_1\zeta^2+a_2\zeta^3+a_3\zeta^4+\cdots$ the first coefficients are
\begin{equation}
g_0=a_1,\qquad g_1=\frac{2(a_2-2a_1^2)}{a_1},\qquad
g_2=-\frac{4(a_2-2a_1^2)(a_2-a_1^2)}{a_1^3}+\frac{3a_3-14a_1a_2+11a_1^3}{a_1^2}.
\end{equation}
Hence, in any dimension, odd powers $m$ are absent if and only if $g_m=0$ for odd $m$; for instance $C_1^{(d)}=0$ iff $a_2=2a_1^2$, independently of $d$. Every statement about integer versus half-integer powers of the stress-tensor invariants that we proved for $d=2$ is therefore true for the whole family.

\subsection{Example: generalised Born--Infeld in $d$ dimensions}

For the generalised Born--Infeld family, $\mathcal F_2=(1-\rho^2)/4$ in section~\ref{sec:GBI}, hence $\mathcal G(\sigma)=\tfrac12(1-2\sigma^2)$, and \eqref{Guniv} gives for any $d$
\begin{equation}
\frac{\partial\mathcal L}{\partial\lambda}=\frac{1}{2d}\Big(X-\frac{2}{d}\,Y\Big).
\label{GBId}
\end{equation}
For $d=2$ this is $\tfrac14(X-Y)$. For $d=4$ it is $\tfrac18(X-\tfrac12Y)$, i.e.\ $C_0=1/8$, $C_2=-1/16$, which is the Born--Infeld flow of electrodynamics. Both cases are the same function $\mathcal G$, evaluated at $d=2$ and $d=4$. The coupling is normalised so that $\lambda$ has dimension $-d$.

\subsection{A $d$-explicit form of the flow and the general $\mathcal G$}

The parametric form makes the dimension dependence transparent. Since $X F_d=a r/\lambda=\tau\ell'\,r$ and $\tau\ell'=a/\lambda$, eliminating $\lambda$ through $\zeta=\zeta(r)$ gives
\begin{equation}
 \frac{\partial\mathcal L}{\partial\lambda}=\frac{dX-Y}{d^2}\;\Psi(r),\qquad
r=-\frac{T^\mu{}_\mu}{\sqrt{dX-Y}},\qquad \Psi(r)\equiv\frac{r}{a(\zeta(r))}.
\label{Psid}
\end{equation}
Here $\Psi$ is a function of $r$ determined by $f$ alone, and the same for all $d$; the dimension appears only through $dX-Y$, $d^2$ and the definition of $r$. It is related to the functions above by $\Psi=(1+r^2)\,\mathcal G=d\,(1+r^2)\,\mathcal F_d$. Equation \eqref{Psid} is therefore the ``one-line'' version of \eqref{flowd}: the four-dimensional flow is obtained from the two-dimensional one by replacing $2X-Y\to4X-Y$ and $\tfrac14\to\tfrac1{16}$, with the same $\Psi$.

For the Russo--Townsend description of section~\ref{sec:RT} the $d$-dimensional version is $T^\mu{}_\mu=-d\,\Omega$, $X=d[(y\Omega_y)^2+\Omega^2]$, $Y=d^2\Omega^2$ and $\partial_\lambda\mathcal L=\Omega/\lambda$, so the potentials of table~\ref{tab:potentials} are $d$-independent, and the weak-field coefficients are
\begin{equation}
\begin{aligned}
C_0^{(d)}&=\frac1{4d\,\omega_2},\qquad C_1^{(d)}=-\frac{\omega_3}{d^{3/2}\,\omega_2^{2}},\qquad C_2^{(d)}= - \frac{\omega_2^2 - 11 \omega_3^2}{4 d^2 \omega_2^3} -  \frac{3 \omega_4}{d^2 \omega_2^2},\\
C_3^{(d)}&=\frac{12 \omega_3 \omega_4}{d^{5/2} \omega_2^3} -  \frac{10 \omega_3^3 -  \omega_2^2 (\omega_3 - 16 \omega_5)}{2 d^{5/2} \omega_2^4}
\end{aligned}
\end{equation}
For $d=2$ these are \eqref{C01omega}.

\subsection{The $q$-deformed models in $d$ dimensions}

Take $f_q=1-s^q$, $s=1-\zeta/q$, as in \eqref{ellq}, with $[\lambda]=-d$. The functions $a_q$, $b_q$, $r_q(s)$, the inversion \eqref{sqdef} and the branch table of section~\ref{sec:q} are unchanged, because they involve $\zeta$ only. In \eqref{Psid},
\begin{equation}
\Psi_q(r)=\frac{r\,s_q^{\,1-q}}{q\,(1-s_q)},\qquad s_q^{\,1-q}=q+(1-q)s_q-q(1-s_q)\,r,
\end{equation}
so that
\begin{equation}
\frac{\partial\mathcal L_q}{\partial\lambda}=\frac{dX-Y}{d^2}\,\frac{r\,s_q^{\,1-q}}{q\,(1-s_q)}\ .
\label{qd}
\end{equation}
In the variable $\rho=\sqrt{Y/X}$, using $r/(1+r^2)=\rho\sqrt{d-\rho^2}/d$,
\begin{equation}
\mathcal F_q^{(d)}(\rho)=\frac{\rho\sqrt{d-\rho^2}}{d^2\,q}\,\frac{s_q^{\,1-q}}{1-s_q},\qquad
r=\frac{\rho}{\sqrt{d-\rho^2}},
\end{equation}
with domains
\begin{equation}
0<\rho<\sqrt{d/2}\ \ (0<q<1),\qquad 0<\rho<\frac{\sqrt d\,(q-1)}{\sqrt{q^2+(q-1)^2}}\ \ (q>1).
\end{equation}
The bounds $0<r<1$ and $0<r<(q-1)/q$ are $d$-independent, and the signs of $\lambda$ (positive for $q<1$, negative for $q>1$) do not change. The coefficients are $C_m^{(d)}=(2/d)^{1+m/2}C_m(q)$ with $C_m(q)$ from \eqref{Cq}:
\begin{equation}
C_0^{(d)}=\frac{1-q}{2d\,q},\quad
C_1^{(d)}=\frac{2(2q-1)}{3\,q\,d^{3/2}},\quad
C_2^{(d)}=-\frac{(q+1)(8q-7)}{18\,q(q-1)\,d^{2}},\ \dots
\end{equation}
The singular points of table~\ref{tab:qsing} are located at the same $r$ and the same fields $(\tau,\ell',p/\mathcal E)$ in every dimension; only the values $\rho_\ast=\sqrt{d}\,r_\ast/\sqrt{1+r_\ast^2}$ of the invariant ratio depend on $d$. The special values are
\begin{itemize}
\item $q=\tfrac12$: $\Psi=\tfrac12(1-r^2)$, giving the Born--Infeld flow \eqref{GBId};
\item $q=1$: $r\equiv0$, ModMax, $\lambda$ drops out;
\item $q=\tfrac34$: $C^{(d)}_m=(2/d)^{1+m/2}\{\tfrac1{12},\tfrac{\sqrt2}9,-\tfrac7{54},-\tfrac{23\sqrt2}{324},\tfrac7{243},\dots\}$;
\item $q=\tfrac32$: the no-$\tau$-maximum model below, with $\lambda_q=-\tfrac32\lambda_{\rm NM}$.
\end{itemize}

\subsection{The no-$\tau$-maximum model in $d$ dimensions}

For $\ell_{\rm NM}=\frac2{3\lambda}[(1+\zeta)^{3/2}-1]$ we have $a=s(s^2-1)$, $b=-\tfrac13(s-1)^2(s+2)$, $s=\sqrt{1+\zeta}$, and $r(s)=\tfrac13-\tfrac2{3s(s+1)}\in[0,\tfrac13)$, $s(r)=\tfrac12\big[\sqrt{3(3-r)/(1-3r)}-1\big]$, all independent of $d$. Then
\begin{equation}
 \frac{\partial\mathcal L_{\rm NM}}{\partial\lambda}=\frac{dX-Y}{24\,d^2}\,(1-3r)^2\Bigg[3+\sqrt{\frac{3(3-r)}{1-3r}}\Bigg],
\qquad r=-\frac{T^\mu{}_\mu}{\sqrt{dX-Y}},
\label{NMd}
\end{equation}
and, in terms of $\rho$,
\begin{equation}
\mathcal F_{\rm NM}^{(d)}=\frac{(1-3r)^2}{24\,d\,(1+r^2)}\Bigg[3+\sqrt{\frac{3(3-r)}{1-3r}}\Bigg]
=\frac1{4d}-\frac4{3d^{3/2}}\rho+\frac{25}{18d^2}\rho^2+\frac{22}{27d^{5/2}}\rho^3+\frac{16}{81d^3}\rho^4+\dots,
\end{equation}
valid for $0\le\rho<\sqrt{d/10}$. Near the endpoint $r\to\tfrac13$, $\mathcal F^{(d)}_{\rm NM}\simeq\frac{3\sqrt2}{40\,d}(1-3r)^{3/2}$, a branch point of the invariants; it again reflects $s,\tau\to\infty$ and the absence of stress tensors with $p/\mathcal E<\tfrac12$. For $d=2$ \eqref{NMd} reduces to the flow of section~\ref{sec:NM}, and for $d=4$ it is its four-dimensional counterpart.

\subsection{The logarithmic model in $d$ dimensions}

For $\ell_{\log}=-\lambda^{-1}\ln(1-\zeta)$ we have $a=(1-s)/s$, $b=1-1/s-\ln s$, $s=1-\zeta$, $r=1+s\ln s/(1-s)$ and $s(r)=(r-1)/W_{-1}\big((r-1)e^{r-1}\big)$. Hence $\Psi_{\log}=r\,s/(1-s)=r(r-1)/D(r)$ with $D(r)=W_{-1}((r-1)e^{r-1})-(r-1)$, and
\begin{equation}
 \frac{\partial\mathcal L_{\log}}{\partial\lambda}=\frac{dX-Y}{d^2}\,\frac{r(r-1)}{W_{-1}\big((r-1)e^{r-1}\big)-(r-1)}\,\qquad
r=-\frac{T^\mu{}_\mu}{\sqrt{dX-Y}}.
\label{logd}
\end{equation}
Equivalently, since $(dX-Y)/d^2=(\tau\ell')^2$ and $r-1=-\ell/(\tau\ell')$, \eqref{logd} is \eqref{logcompact} for every $d$. In $\rho$,
\begin{equation}
\mathcal F^{(d)}_{\log}=\frac{r\,s(r)}{d\,(1-s(r))(1+r^2)}=\frac1{2d}-\frac{2}{3d^{3/2}}\rho-\frac{7}{18d^2}\rho^2+\frac{49}{135d^{5/2}}\rho^3+\frac{4}{405d^3}\rho^4+\dots,
\end{equation}
for $0<\rho<\sqrt{d/2}$ ($0<r<1$). The analysis of the singular points in section~\ref{sec:log} carries over verbatim: the removable $0/0$ at $r=0$ (limit $\tfrac12(\tau\ell')^2$), the smooth window $0<r<1$ where $D<0$, and the logarithmic branch point $\partial_\lambda\mathcal L_{\log}\simeq(\tau\ell')^2(1-r)/|\ln(1-r)|$ at the critical current $v\to e^{-\gamma}/\lambda$ are statements about $r$ and $\zeta$, and so are independent of $d$.

\subsection{Summary for arbitrary $d$ and Remarks}

\begin{table*}[t]
\centering\small
\begin{tabular}{@{}lllll@{}}
\toprule
Model & $\Psi(r)$ in \eqref{Psid} & $r$-range & $\rho$-range & $C_0^{(d)},\ C_1^{(d)}$\\
\midrule
GBI $(q=\tfrac12)$ & $\tfrac12(1-r^2)$ & $[0,1)$ & $[0,\sqrt{d/2})$ & $\tfrac1{2d},\ 0$\\
$q$-deformed & $\dfrac{r\,s_q^{1-q}}{q(1-s_q)}$ & $[0,1)$, $q<1$; $[0,\tfrac{q-1}q)$, $q>1$ & $[0,\sqrt{d/2})$; $[0,\tfrac{\sqrt d(q-1)}{\sqrt{q^2+(q-1)^2}})$ & $\tfrac{1-q}{2dq},\ \tfrac{2(2q-1)}{3qd^{3/2}}$\\
No-$\tau$-max & $\tfrac{(1-3r)^2}{24}\Big[3+\sqrt{\tfrac{3(3-r)}{1-3r}}\Big]$ & $[0,\tfrac13)$ & $[0,\sqrt{d/10})$ & $\tfrac1{4d},\ -\tfrac4{3d^{3/2}}$\\
Logarithmic & $\dfrac{r(r-1)}{W_{-1}((r-1)e^{r-1})-(r-1)}$ & $[0,1)$ & $[0,\sqrt{d/2})$ & $\tfrac1{2d},\ -\tfrac2{3d^{3/2}}$\\
\bottomrule
\end{tabular}
\caption{The one-scale flows $\partial_\lambda\mathcal L=\frac{dX-Y}{d^2}\Psi(r)$ in $d$ dimensions, $r=-T^\mu{}_\mu/\sqrt{dX-Y}$, $\rho^2=Y/X=dr^2/(1+r^2)$. The columns for $\Psi$ and the $r$-range are independent of $d$, and $C^{(d)}_m=(2/d)^{1+m/2}C^{\rm 2D}_m$.}
\label{tab:anyd}
\end{table*}


(i) The root flow in \eqref{flowd} is also universal, $\partial_\gamma\mathcal L=\sqrt{dX-Y}/d$, and the commutativity proof of section~\ref{sec:comm} holds without change in any $d$, since it uses only the Courant--Hilbert form through $\ell_{\lambda\gamma}-N\ell'_\lambda\ell'_\gamma$ and is blind to the dimension.
(ii) The class \eqref{Tgend} requires $d$ to be even if $S_{\mu\nu}$ has the $\pm1$ eigenvalues with equal multiplicity. Natural candidates in higher even $d$ are self-dual $(d/2-1)$-form theories~\cite{Bandos:2020hgy,Avetisyan:2022zza,Ferko:2024zth,Russo:2025wph}, for which $T\bar T$-like flows are known, for instance in $d=6$~\cite{Ferko:2024zth}. For odd $d$, \eqref{flowd}--\eqref{Guniv} should be read as the formal continuation in $d$, and we do not use it.
(iii) Equations \eqref{flowd} and \eqref{Guniv} make explicit that the two-dimensional integrable sigma models and four-dimensional self-dual electrodynamics are two realisations of one Courant--Hilbert structure.

\section{Commutativity with the root-$T\bar T$ flow}\label{sec:comm}

A one-scale family $\mathcal L(\lambda,\gamma)$ is driven by two autonomous stress-tensor flows,
\begin{equation}
\partial_\gamma\mathcal L=\mathcal R_\gamma\equiv\tau\ell'=\tfrac12\sqrt{2X-Y},\qquad
\partial_\lambda\mathcal L=\mathcal O_\lambda\equiv X\,\mathcal F\!\big(\sqrt{Y/X}\big).
\end{equation}
Neither operator depends explicitly on $\lambda$ or $\gamma$.

\paragraph{Proposition.} For every one-scale family, with all derivatives at fixed $(u,v)$,
\begin{equation}
\partial_\lambda\big(\mathcal R_\gamma[\mathcal L]\big)=\partial_\gamma\big(\mathcal O_\lambda[\mathcal L]\big)=\partial_\lambda\partial_\gamma\mathcal L .
\label{commute}
\end{equation}
Equivalently, the two-parameter family is reached from the free (PCM) point by applying $\mathcal O_\lambda$ and $\mathcal R_\gamma$ in either order
(figure~\ref{fig:commute}).

\emph{Proof.} Write $\ell_\alpha=\partial_\alpha\ell|_\tau$, $\ell'_\alpha=\partial_\alpha\ell'|_\tau$ for $\alpha\in\{\lambda,\gamma\}$. By Lemma~1,
$\partial_\alpha\mathcal L|_{u,v}=\ell_\alpha(\tau)$. Varying the constraint $\tau=v+u/\ell'^2$ at fixed $u,v$,
$\partial_\alpha\tau|_{u,v}=-N\,\ell'_\alpha$ with $N=\dfrac{2u/\ell'^3}{1+2u\ell''/\ell'^3}$. Hence
\begin{equation}
\partial_\beta\partial_\alpha\mathcal L=\ell_{\alpha\beta}-N\,\ell'_\alpha\ell'_\beta ,
\end{equation}
which is symmetric in $(\alpha,\beta)$. Because $\partial_\gamma\mathcal L=\mathcal R_\gamma[\mathcal L]$ and $\partial_\lambda\mathcal L=\mathcal O_\lambda[\mathcal L]$ hold
identically in $(\lambda,\gamma)$, differentiating the first in $\lambda$ and the second in $\gamma$ gives \eqref{commute}. $\square$

For the one-scale form, $\ell_{\lambda\gamma}=\zeta^2f''/\lambda^2$, $\ell'_\lambda=e^\gamma\zeta f''/\lambda$, $\ell'_\gamma=e^\gamma(f'+\zeta f'')$ and $\ell''=e^{2\gamma}\lambda f''$.
We verified \eqref{commute} numerically for the GBI, $q=\tfrac34$, no-$\tau$-maximum and logarithmic models to $\sim10^{-9}$. The statement and proof are dimension independent
and hold verbatim for the four-dimensional theories~\cite{Pezhman:2026dsr,Pezhman:2026sby,Pezhman:2026gbv}, and they are the exact counterpart of the perturbative commutativity of \cite{Babaei-Aghbolagh:2024uqp,Babaei-Aghbolagh:2024hti}.

\begin{figure}[t]
\centering
\begin{tikzcd}[row sep=huge,column sep=huge]
		\mathcal{L}_{PCM} \arrow[r, blue, "{\cal O}_{\lambda}" blue] \arrow[d,red,"{\cal R}_{\gamma}" red]
		&|[blue]| \mathcal{L}(\lambda, 0) \arrow[d,red, "{\cal R}_{\gamma}" red] \\
		|[red]|\mathcal{L}_{MM}( \gamma) \arrow[r, blue, "{\cal O}_{\lambda}" blue]
		&|[red!50!blue]|  \mathcal{L}(\lambda, \gamma)
	\end{tikzcd}
\caption{Commutative diagram for the irrelevant operator $\mathcal O_\lambda=X\mathcal F(Y/X)$ and the root operator $\mathcal R_\gamma=\tfrac12\sqrt{2X-Y}$.
For the GBI family, $\mathcal L(\lambda,0)$ is the $T\bar T$-deformed PCM, $\mathcal L_{\rm MM}(\gamma)$ the root-$T\bar T$-deformed PCM (2D ModMax) and $\mathcal L(\lambda,\gamma)$ the general deformed model.}
\label{fig:commute}
\end{figure}

\section{Discussion and outlook}\label{sec:conclusions}
We determined the exact $T\bar T$-like operator of two-dimensional integrable sigma models whose Lagrangian solves $\mathcal L_u\mathcal L_v=-1$ and has a one-scale Courant--Hilbert function $\ell=\lambda^{-1}f(e^\gamma\lambda\tau)$. The stress tensor is controlled by $\tau\ell'$ and $\ell-\tau\ell'$, so the invariants $X$ and $Y$ are functions of one variable; eliminating it gives the autonomous flow $\partial_\lambda\mathcal L=X\,\mathcal F(\sqrt{Y/X})$, together with the universal root-$T\bar T$ flow $\partial_\gamma\mathcal L=\tfrac12\sqrt{2X-Y}$, which commutes with it (section~\ref{sec:comm}). The Russo--Townsend potential is the Legendre transform of the same data and plays the role of the interaction function of the auxiliary field sigma model; since the Lax connection of the PCM is common to all members, every flow considered preserves classical integrability. Table~\ref{tab:summary} lists the four exact examples. Section~\ref{sec:anyd} shows that the same equations, written in $d$ dimensions, give the four-dimensional electrodynamics flow at $d=4$, with all dimension dependence in the prefactor $(dX-Y)/d^2$ and in $r=-T^\mu{}_\mu/\sqrt{dX-Y}$.

\begin{table*}[t]
\centering\small
\begin{tabular}{@{}lllll@{}}
\toprule
Model & range of $\rho$ & $C_0$ & $C_1$ & flow\\
\midrule
GBI ($q=\frac12$) & $[0,1]$ & $\frac14$ & $0$ & $\frac14(X-Y)$, polynomial\\
$q$-deformed, $q<1$ & $[0,1)$ & $\frac{1-q}{4q}$ & $\frac{\sqrt2(2q-1)}{6q}$ & implicit, $s_q$ from \eqref{sqdef}\\
$q>1$ ($\lambda<0$) & $[0,\frac{\sqrt2(q-1)}{\sqrt{q^2+(q-1)^2}})$ & $\frac{1-q}{4q}$ & $\frac{\sqrt2(2q-1)}{6q}$ & implicit\\
no-$\tau$-max.\ ($q=\frac32$) & $[0,\frac1{\sqrt5})$ & $\frac18$ & $-\frac{\sqrt2}3$ & closed form, algebraic\\
logarithmic ($q\to0$) & $[0,1)$ & $\frac14$ & $-\frac{\sqrt2}6$ & Lambert $W_{-1}$\\
\bottomrule
\end{tabular}
\caption{Summary of the exact $T\bar T$-like flows $\partial_\lambda\mathcal L=X\mathcal F(\rho)=\sum_mC_mY^{m/2}X^{1-m/2}$. For $q>1$ and the no-$\tau$-maximum model the $C_m$ refer to the coupling $\lambda_{\rm NM}>0$ in which the model is causal.}
\label{tab:summary}
\end{table*}

Only the GBI model truncates to integer powers of $X$ and $Y$; all other members contain half-integer powers, equivalently a non-analyticity in $T^\mu{}_\mu$ at the conformal point which is, however, a
coordinate artefact of $r\ge0$. The endpoints of the range of $\rho$ have a direct meaning: for $q<1$ and for the logarithmic model they are the critical value of $v$ where $\ell'\to\infty$ (a power-law branch
point for $q$-models, logarithmic for the log model); for $q>1$ and for the no-$\tau$-maximum model they correspond to $\tau\to\infty$ with $p/\mathcal E\to1/(2q-1)$.

The construction applies to one-scale solutions, $\mathcal F$ is family-specific rather than universal, it is single-valued only where $\zeta\mapsto\rho$ is invertible
(checked for the examples, not proved for general $f$), and the limit $q\to1$ is degenerate since $r\equiv0$.

Natural directions are the inverse problem (reconstructing $f$ from a given $\mathcal F$, with causality constraints), the classification of one-scale solutions, the deformation of the Maillet bracket~\cite{Maillet:1985ek,Maillet:1985ec} and
Yangian charges along these flows~\cite{Bielli:2026ggs}, the fate of the Lax connection under $T\bar T$-like flows~\cite{Chen:2021aid}, extension beyond one-scale families and to cosets~\cite{Cesaro:2024ipq}, Yang--Baxter models~\cite{Bielli:2024fnp,Fukushima:2026gan} and $\sigma$-models with Wess--Zumino term, and the quantum meaning of the non-polynomial operators. The close relation
to the four-dimensional results~\cite{Babaei-Aghbolagh:2025cni,Pezhman:2026dsr,Pezhman:2026sby,Pezhman:2026gbv}, where $C_m^{\rm 4D}=2^{-1-m/2}C_m^{\rm 2D}$, suggests a uniform treatment of self-dual systems in any dimension; this is made explicit in section~\ref{sec:anyd}. A natural next step is to identify, in each even $d$, the actual field theories realising the stress-tensor structure assumed there beyond $d=2,4$ (for instance self-dual $(d/2-1)$-form theories), and to test whether the $d$-independent function $\Psi$ continues to control their flows.

\begin{acknowledgments}
The author is grateful to Komeil Babaei Velni, Christian Ferko, Song He, Stefano Negro, Zahra Pezhman, Jorge G. Russo, Dmitri Sorokin, and Roberto Tateo for their interest in this work and the fruitful discussions that followed. This work was conducted as part of the PostDoc Program on {\it Exploring TT-bar Deformations: Quantum Field Theory and Applications}, sponsored by Ningbo University. 
\end{acknowledgments}

\bibliographystyle{JHEP}
\bibliography{Ref_2D}
\end{document}